\documentclass{raa} 
\usepackage{graphicx,times}
\usepackage{natbib}
\usepackage{amssymb,amsmath}
\usepackage{hyperref}
\usepackage{soul}
\usepackage{epsfig}
\usepackage{multirow}
\usepackage{enumerate}
\usepackage{amsmath}
\usepackage{float}
\usepackage[section]{placeins}
\usepackage{url}
\usepackage{color}
\bibpunct{(}{)}{;}{a}{}{,}
\newcommand{\Rmnum}[1]{\uppercase\expandafter{\romannumeral #1}}
\newcommand{\tabincell}[2]{\begin{tabular}{@{}#1@{}}#2\end{tabular}}

\begin{document}

\title{Statistical properties of fast radio bursts elucidate their origins: magnetars are favoured over gamma-ray bursts}

\volnopage{Vol.0 (20xx) No.0, 000--000}     
\setcounter{page}{1}

\author{Xiang-Han Cui\inst{1,2}, Cheng-Min Zhang\inst{1,2,3}, Shuang-Qiang Wang\inst{4}, Jian-Wei Zhang\inst{5}, Di Li\inst{1,2,6}, Bo Peng\inst{1,2,7}, Wei-Wei Zhu\inst{1,2},  Richard  Strom\inst{8,9}, Na Wang\inst{4}, Qingdong Wu\inst{4}, Chang-Qing Ye\inst{10},  De-Hua Wang\inst{11}, Yi-Yan Yang\inst{11}, Zhen-Qi Diao\inst{11}}

\institute{
CAS key Laboratory of FAST, National Astronomical Observatories, Chinese Academy of Sciences, Beijing 100101, China \it{zhangcm@bao.ac.cn(CMZ)}
\and School of Astronomy and Space Science, University of Chinese Academy of Sciences, Beijing 100049, China
\and School of Physical Sciences, University of Chinese Academy of Sciences, Beijing 100049, China
\and Xinjiang Astronomical Observatory, Chinese Academy of Sciences, Urumqi, Xinjiang 830011, China
\and Department of Astronomy, Beijing Normal University, Beijing, 100875, China
\and NAOC-UKZN Computational Astrophysics Centre, University of KwaZulu-Natal, Durban 4000, South Africa
\and Key Laboratory of Radio Astronomy, Chinese Academy of Sciences, Beijing 100101, China
\and Netherlands Institute for Radio Astronomy (ASTRON), Postbus 2, 7990 AA Dwingeloo, the Netherlands
\and Astronomical Institute ‘Anton Pannekoek’, Faculty of Science, University of Amsterdam, 1090 GE Amsterdam, the Netherlands
\and TianQin Research Center for Gravitational Physics, Sun Yat-sen University, Zhuhai 519082, China
\and School of Physics and Electronic Sciences, Guizhou Education University, Guiyang 550018, China
}


\abstract{
Fast radio bursts (FRBs) are extremely strong radio flares
lasting several milliseconds, most of which come from
unidentified objects at a cosmological distance. They can be apparently repeating or not.
In this paper, we analyzed 18 repeaters and 12 non-repeating FRBs observed in the frequency
bands of 400-800 MHz from CHIME. We investigated the distributions
of FRB isotropic-equivalent radio luminosity, considering the K correction.
Statistically, the luminosity distribution can be better fitted by Gaussian form than by power-law.
Based on the above results, together
with the observed FRB event rate, pulse duration, and radio luminosity,
FRB origin models are evaluated and constrained such that the
gamma-ray bursts (GRBs) may be excluded for the
non-repeaters while magnetars or neutron stars (NSs) 
emitting the supergiant pulses are preferred for the repeaters.
We also found the necessity of a small
FRB emission beaming solid angle (about 0.1 sr) from magnetars that should be
considered, and/or the FRB association with soft gamma-ray
repeaters (SGRs) may lie at a low probability of about 10\%.
Finally, we discussed the uncertainty of FRB luminosity
caused by the estimation of the distance that is inferred by the
simple relation between the redshift and dispersion measure (DM).
}

\titlerunning{Statistical Properties of Fast Radio Bursts}
\authorrunning{Cui et al.}

\maketitle

\section{Introduction}\label{1}

Fast radio bursts (FRBs) are very strong radio emissions in  a couple of milliseconds, which are mostly confirmed to be from cosmic distances.
The FRB phenomenon  was firstly noticed and reported  in 2007 \citep{Lorimer07},
and the first  non-Parkes FRB event, FRB 121102,  was observed by the Arecibo telescope  in 2014 \citep{Spitler14}, which
is also the first confirmed  repeating FRB with the localized host galaxy\citep{Spitler16, Chatterjee17}.
With the completion of the advanced radio instrumentations like Canadian Hydrogen Intensity Mapping Experiment (CHIME) \citep{CHIME19a,CHIME19b,CHIME20a,CHIME20b}, Australian Square Kilometre Array Pathfinder (ASKAP) \citep{Shannon18, Kumar19} and Five-hundred-meter Aperture Spherical radio Telescope (FAST) \citep{Li18, Zhu20, Luo20b, Lin20}, the number of  FRBs has dramatically  increased \citep{Petroff16}.
Under the continuous endeavour  on FRB searching    over a decade \citep{Lorimer18},
up to now there are 22 repeaters and 107 apparently non-repeaters published
with the event rate of $10^{3-4}\,day^{-1}\,sky^{-1}$ \citep{Lorimer07, Thornton13,Spitler14,Kulkarni14, Keane15,Rane16,Oppermann16,Champion16,Scholz16,Lawrence17,Patel18,Connor19}.
Recently, FRB-like signals (named as FRB 200428) from a  galactic magnetar that is identified as a soft gamma-ray repeater (SGR) 1935+2154 have been detected by STARE2 \citep{Bochenek20} and CHIME \citep{CHIME20b}, strongly supporting
the idea that the magnetar is the promise candidate for FRB origin.
Cosmological FRBs are still remained as the unsettled questions \citep{Kulkarni14,Keane18,Pen18,Cordes19}.
At present, there are lots of  theoretical models proposed to explain the origin and radiation mechanism of FRBs,
most of which are mainly centered at the ideas  borrowed from the  pulsars \citep{Cordes16} and gamma-ray bursts (GRBs)
 \citep{Zhang14}.

To constrain FRB models, we performed statistical tests on the intrinsic duration and isotropic-equivalent radio luminosity,
concluding that the repeaters and apparently non-repeaters should have different origins or physical processes \citep{Cui20}.
This indicates that at least two different models are needed to explain these two classes of FRBs.
Since FRB phenomena were observed in the magnetar   SGR 1935+2154 (FRB 200428), the  FRB-SGR association is confirmed \citep{Bochenek20, CHIME20b,  Lin20, Li20},  although its estimated radio emission luminosity  is about 4 orders of magnitudes lower  than that of the observed cosmological FRBs.
Meanwhile, magnetars can act as the central engines of both pulsar-like and GRB-like models (see the review of \cite{Zhang20}),
but the other alternative schemes cannot be completely ruled out.
Therefore, it is necessary to investigate  more aspects of FRB statistical properties to constrain FRB models.

FRB luminosity distribution has  been studied by several researchers  \citep{Kumar17, Li17, Luo18, Hashimoto20, Luo20a}, the power-law type with various indices is usually preferred  \citep{Zhang20}.
In the work from \cite{Li17}, they analyzed observed fluence of 16 non-repeaters, deriving a power-law type for intensity distribution function.
Meanwhile \cite{Luo20a} assumed the Schechter function as a likelihood function in Bayesian method, while \cite{Hashimoto20} applied a simple $\rm V_{max}$ method without any presupposition functional shape in their analysis.
As expected, the different fitting functions for FRB luminosity distributions  should be rooted in the particular physical origins or radiation processes of FRBs, by which the FRB models could be constrained.

In this paper, we divide the isotropic-equivalent radio luminosity data  from CHIME with K correction into two sample sets according to the repeatability \citep{Petroff19, Cui20}.
Then,  we compare the  two different types of fitting functions, the Gaussian type and power-law type, respectively, based on the goodness of fittings.
Furthermore, combining the statistical properties of FRB duration and luminosity, as well as the event rate of FRBs,
we obtain that models associated with GRBs for the non-repeater are ruled out,
while the repeating FRBs favor magnetars or supergiant pulses as their origins.
In addition, if we consider the low probability of SGRs to exhibit FRBs (e.g.,  $\sim 10\%$), and the
small beaming angle of FRB emission  (e.g., 0.1 sr), then the
tension between the birth rate of magnetars and  the event rate of FRBs can be settled down.

The structure of our paper is as follows.
In Section 2, we describe the data selection of two samples.
In Section 3, we fit the FRB isotropic-equivalent radio luminosity distribution by both power-law and Gaussian functions.
In Section 4, we discuss the constraints on the different FRB models and evaluate the possibility of producing repeating FRBs by magnetars.
Finally, in Section 5, we summarize our results.

\section{Selection of Observation Data }\label{2}

To limit the uncertainties of the FRB flux density resulted
from the different types of radio telescopes,  we select the FRB
data only from CHIME (12 non-repeaters and 18 repeaters), detected
at frequency band of 400-800 MHz \citep{CHIME19a,CHIME19b}, since
the thresholds and beam patterns of FRBs are related to
the  features of the radio telescopes
\citep{Caleb16}.

In  previous investigation  about the FRB luminosity function,  a Schechter type is selected as described in the following  Eq. (1),
 which is usually considered as a likelihood function in Bayesian method \citep{Luo18, Luo20a},
\begin{equation}
\begin{split}
\phi(\log L) \mathrm{d} \log L=\rho\left(\frac{L}{L_{max}}\right)^{\alpha+1} {\rm e}^{-\frac{L}{L_{max}}} \mathrm{d} \log L ,
\end{split}
\end{equation}
where $\rho$ is a normalization coefficient, $L$ is the radio luminosity, $L_{max}$ is the upper limit of FRB luminosity (hereafter we take it as
 the maximum value of FRBs) and $\alpha$ is the power-law index.
The form of the likelihood function has affect on the final luminosity function.
Inappropriate selection of likelihood function may cause errors and misunderstandings when we attempt to explain these observations.
Hence, it is necessary to check the selected type of the likelihood function.
The Schechter type and power-law are different in terms of the function type, but Schechter function is still a special
case of power-law type, containing the exponential cutoff at the high end.
Here we employ the  Schechter function to represent the power-law type.
Without considering the scattering \citep{Lorimer13},  the isotropic-equivalent radio luminosity (hereafter referred to as radio luminosity)
of an FRB is defined by Eq. (2),
\begin{equation}
\begin{split}
L_{\rm{iso}}\sim S_{\nu_c}D_{L}^2,
\end{split}
\end{equation}
where $S_{\nu_c}$ is the flux density at central frequency $\nu_c$ and $D_{L}$ is the FRB luminosity distance.

Whether the repeaters and apparently non-repeaters
originated from the same population is still unclear, there are
several works that have discussed this issue.  In the views of
\cite{Fonseca20} and \cite{Cui20}, they believe that two
types of FRBs follow different origins. However,  the
simulation from \cite{Gardenier20}  show that the observed FRBs
in the sky can be accounted by a single population with varying
repetition rate. The above works share similar  views   that
the  repeaters and apparently non-repeaters may come
from  different physical processes or traits. So in the
following analysis, based on the occurrence property (repeaters
and apparently non-repeaters), we divide the CHIEM FRB data into
two  sample groups. Group one is the published repeating FRB
data \citep{CHIME19a,CHIME19b,Fonseca20}, and group two is the
announced non-repeating FRB sample, corresponding to the database
from CHIME and FRB Catalogue (FRBCAT)
\footnote{\url{http://www.frbcat.org/}} \citep{Petroff16}. Some
repeating FRBs were previously thought to be the non-repeater ones
\citep{Kumar19}, however it is difficult to foresee whether all
non-repeaters are bound to burst again in the future. So, based
on the current situation, we presume  the "apparently"
non-repeating sources as the real non-repeaters. Considering that the
repeaters contain multiple bursts, we take their mean value as a
statistical variable for each repeater.

Up to now,  the distances of 13 FRBs are measured directly by the
redshift of their host galaxy\footnote{\url{http://frbhosts.org/}},
while the  distances of the other FRBs are all estimated by their
dispersion measures (DMs). In the FRBCAT, \cite{Petroff16}
obtained the maximum redshift by DM and inferred the FRB
luminosity distance. The $\rm {\Lambda CDM}$
cosmological parameters that they employed are listed below:
$\rm{H_0=69.6\,km\,s^{-1}\,Mpc^{-1}}$, $\rm{\Omega_M=0.286}$ and
$\rm{\Omega_{vac}=0.714}$, where $\rm{H_0}$ is the Hubble
constant, and $\rm{\Omega_M}$ and $\rm{\Omega_{vac}}$ are the matter
and dark energy fraction in the universe, respectively.
In our analysis, the FRB luminosity distance is directly taken
from the FRBCAT, since  the  estimated distances by
\cite{Luo18} share the similar results to those of FRBCAT  (see
Appendix B).
As noted, the distances of FRBs in the above two samples are
corresponding to the  redshift (z) from 0.05 to 2.1, so the FRB
luminosity needs a K correction, which is described in Eq.(3) as
below  \citep{Hogg02, Xiao21}. 
\begin{equation}
\begin{split}
L_{\rm{k}}\sim S_{\nu_c}{D_{L}^2\over (1+z)}.
\end{split}
\end{equation}

\section{Luminosity distribution of FRBs}\label{3}
Here, for the two groups of FRB luminosity data as mentioned above,
we fit their distributions by  the  two types of functions, a power-law type and Gaussian type, respectively.
Then we compare the goodness of  fitting results to test which form of function is more appropriate.
The histograms of repeaters and fitted curves of the different types  are shown in Figure 1.
The goodness of power-law and Gaussian form are 0.013 and 0.927, respectively, for the
repeater samples.
Figure 2 shows the histogram and fitted curves of non-repeaters,
and the goodness of power-law and Gaussian form are 0.006 and 0.734, respectively.

What we have to clarify here is that although the goodness of Gaussian form is much better than power-law type,
we still can not draw a conclusion that the Gaussian is the best form for luminosity distribution.
But at least, the fittings of power-law type is not suitable enough for the luminosity distribution from CHIME data, and Gaussian type is better than power-law type.

\begin{figure}
\centering
\includegraphics[width=10cm]{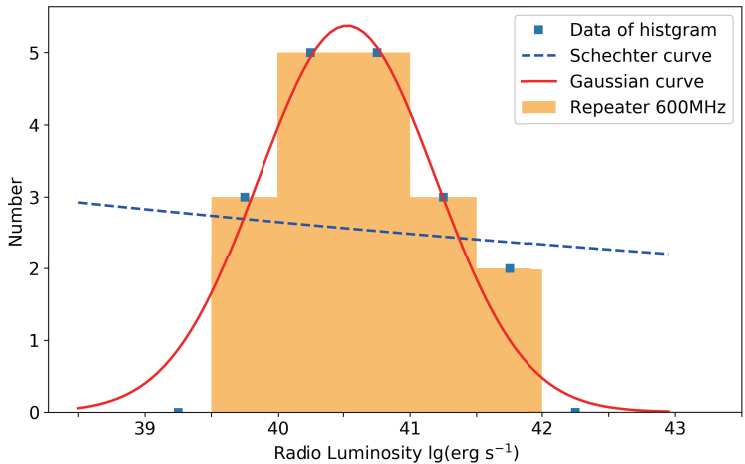}
\caption{Histogram and different fitted curves of repeaters at 600MHz.
The solid line is the curve of Gaussian form and the dashed line is the curve of power-law form.
The square dots are the data points.}
\label{fig2}
\end{figure}

\begin{figure}
\centering
\includegraphics[width=10cm]{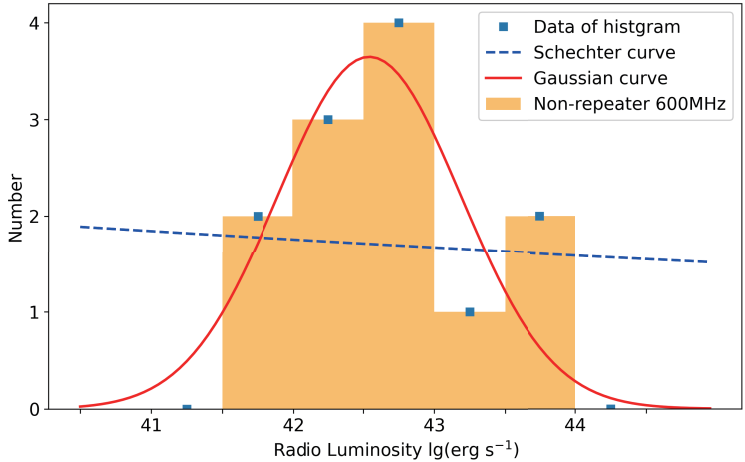}
\caption{Histogram and different fitted curves of non-repeaters at 600MHz.
The solid line is the curve of Gaussian form and the dashed line is the curve of power-law type.
The square dots are the data points.}
\label{fig4}
\end{figure}

\section{Results and Discussions}\label{4}
For the radio luminosity distributions of FRBs in two sample groups,
they are more likely to favor the Gaussian type rather than power-law type (Schechter function),
which can restrict the FRB  models.
The  burst mechanism or radiation process of FRBs  prefer  the sources that emit the radio flashes  to follow
the  Gaussian distribution of luminosity. This fact should rule out or favor some procedures of FRB origins, as described below.
\begin{table*}
\centering
\caption{Summary of FRB models.}
\resizebox{\textwidth}{!}{
\begin{tabular}{@{}lcccccc@{}}
\hline
\hline
\noalign{\smallskip}
\bf Sources &\bf Duration &\tabincell{c}{\bf FRB luminosity\\$erg\,s^{-1}$} &\tabincell{c}{\bf Gaussian\\\bf distribution} &\tabincell{c}{\bf Rate$^a$ \\$Gpc^{-3}\,yr^{-1}$}& \bf Possibility$^b$ & \bf Ref. \\
\hline
\noalign{\smallskip}
\bf FRB observations&&&&&& \\
Non-repeater&$\rm3.35\,ms$$^c$ &$6.2\times 10^{42}$$^d$&yes&\multirow{2}{*}{$\sim 10^4$} & \multirow{2}{*}{-} &\multirow{2}{*}{[1][2][3]} \\
Repeater&$\rm5.10\,ms$$^c$ &$2.6\times 10^{41}$$^d$&yes&&& \\
\hline
{\bf Non-repeaters}&&&&&&\\
LGRB &$\rm >2\,s$&$\le\eta\,10^{54}$$^{e}$&no& $7\times 10^2$ & low & [4][5] \\
SGRB &$\rm <2\,s$&$\sim\eta\,10^{50-52}$&yes& $1.1\times 10^3$ & low & [5][6][7] \\
NS-NS &$\rm \sim ms$&$\sim 10^{45}$&yes& $1.5\times 10^3$ & low & [8][9] \\
NS-WD &$\rm \sim ms$&$\sim 10^{43}$&yes& $\sim 10^4$ & high & [10][11] \\
WD-WD &$\rm \sim ms$& $\sim 10^{42}$&yes& $\sim 10^{4-5}$ & high & [12][13] \\
NS-asteroid &$\rm \sim ms$&$\sim 10^{40}$ &yes& uncertain & uncertain & [14][15] \\
BH-BH &$\rm \sim ms$&$\sim 10^{56}$ &yes& 9-240 & low & [16] \\
BH-NS &$\rm \sim ms$&$\sim 10^{40-41}$&yes&3-20& low & [17]  \\
BH-WD &$\rm \sim ms$&$\sim 10^{40}$&yes&$\sim 10^4$& mediate & [18][19]  \\
\hline
{\bf Repeaters}&&&&&&\\
Magnetar (SGR) &$\rm \sim ms$&$\le 10^{45}$&yes& $\sim 10^{4-7}$$^f$ & high & [20][21][22] \\
NS Supergiant pulse  &$\rm \sim ms$&$\ge 10^{40}$&yes& $\sim 10^5$$^g$ & high & [23][24][25] \\
\hline
\hline
\end{tabular}
}
\label{tab2}
\begin{flushleft}
{\bf Notes.} $^a$ The FRB event rate of different models is expressed  in the unit of $Gpc^{-3}\,yr^{-1}$,
obtained as $\sim  10^4\,Gpc^{-3}\,yr^{-1} \sim 10^4\,day^{-1}\,sky^{-1}$ \citep{Lorimer07, Kulkarni14, Zhang20}
or $\sim 10^{-5}\,yr^{-1}\,galaxy^{-1}$, and the conversion  of the FRB event rate in various units  is described in Appendix A.
Uncertain means that it is hard to estimate the rate on a cosmological scale.
In NS-asteroid model, it is hard to predict the number of asteroid belts or their true density near the
galactic nuclei.

$^b$ Qualitative evaluations of the possibility for various FRB models with 4 levels, low, mediate, high and uncertain.
Low means that in the 4 constrains (duration, FRB energy, luminosity distribution and rate), at least one parameter
is far from the observed constraint of FRBs.
Mediate  means that one parameter   does not meet the observational  constraint of FRBs, but this parameter can be
adjusted according to the model.
High means that in the above 4 conditions, we have no sufficient reasons to rule out that model.
Uncertain means that at least one parameter   cannot be  convinced or ruled out.

$^c$ Mean value of duration.
$^d$ Mean value of isotropic equivalent radio luminosity.

$^e$ $\eta$ stands for the  radio efficiency of FRB transferred from the high energy emissions,
 which is so small as $\sim10^{-4}$, estimated from the magnetar FRB event of
 FRB200428 \citep{Bochenek20}.

$^f$ The FRB event rate of $\sim 10^4\,Gpc^{-3}\,yr^{-1}$ corresponds to  the case  that  the SGR burst
 energy is greater than $\sim 10^{46}\,erg\,s^{-1}$ in $\gamma$-ray band \citep{Ofek07}.
The FRB event rate of $\sim10^6-10^7\,Gpc^{-3}\,yr^{-1}$ is directly inferred  from  the magnetar birth rate of
 $\sim 10^{-2}-10^{-3}\,yr^{-1}\,galaxy^{-1}$ \citep{Kouveliotou98, Gill07, Ferrario08, Gull15, Mereghetti15, Beniamini19},
  where  the rarity factor of SGR-FRB associations and FRB beaming solid angle are not taken into acount.

$^g$ Estimated based on the conditions that only 10\% of
core-collapse supernovae form the  FRB emitters, and each NS ought
to emit at least one  bright supergiant pulse  during its lifetime
with a beaming factor of 0.1.

{\bf Ref.:}
[1]\cite{Lorimer07}, [2]\cite{Kulkarni14}, [3]\cite{Zhang20}
[4]\cite{Chapman07}, [5]\cite{Sun15}, [6]\cite{Berger14}
[7]\cite{Coward12},
[8]\cite{Abbott17}, [9]\cite{Totani13}, [10]\cite{Thompson09}, [11]\cite{Liu18}
[12]\cite{Badenes12}, [13]\cite{Kashiyama13}
[14]\cite{Geng15}, [15]\cite{Dai16}, [16]\cite{Abbott16}, [17]\cite{Mingarelli15},
[18]\cite{Li18}, [19]\cite{Cowperthwaite15}, [20]\cite{Katz16},
[21]\cite{Ofek07}, [22]\cite{Mereghetti15}, [23]\cite{Cordes16}, [24]\cite{Lyutikov16}, [25]\cite{Mu20}.
\end{flushleft}
\end{table*}
\subsection{Ruling out the GRB origin for FRBs}
Usually there are two distinct classes of  GRBs,  long GRBs (LGRBs) and short GRBs (SGRBs), divided by their duration of $\sim$ 1-2 seconds,
which are  assumed to be produced by the collapses of massive stars and  mergers of compact stars,
respectively \citep{Berger14, Abbott17, Jespersen20}.
In the early years of FRB discovery, the origins of FRBs associated with GRBs have been
proposed  \citep{Egorov09, Falcke14, Zhang14, Romero16, Metzger17, Margalit19}.
Based on the statistical properties of FRBs, together with the comparisons of event rates between FRBs and GRBs,
we find that non-repeating  FRBs do not favor the origin models associated with both  LGRBs and SGRBs, and the reasons are listed below and in Table 1.

\Rmnum{1}.
LGRBs  happen  in the final collapses  of massive stars, whose
luminosity distribution is usually described in Schechter function
form \citep{Dwek13, Trenti13, McGuire16}. While the luminosity
distribution of FRBs   prefers the  Gaussian type. 
\Rmnum{2}. The birth  rate of LGRBs is not consistent with that of
FRBs, which are approximately $7\times 10^2 \,Gpc^{-3}\,yr^{-1}$
\citep{Chapman07, Sun15, Luo20a} and $2\times
10^4\,Gpc^{-3}\,yr^{-1}$ \citep{Kulkarni14}, respectively.
\Rmnum{3}. The durations of LGRBs are usually much longer
than 2 seconds, which are far from those of FRBs (milliseconds
timescale). 

SGRBs are considered as the mergers of  compact objects in
binary systems, e.g., double neutron star (DNS) \citep{Berger14,
Abbott17} and black hole (BH)-NS system \citep{D15}. Although the luminosity distribution of DNS mergers can
be taken as a  Gaussian type, the duration and event rate are
not  consistent with those of FRBs. 
Moreover, the birth rate  of SGRBs is about  $1.1\times 10^3
\,Gpc^{-3}\,yr^{-1}$ \citep{Coward12, Sun15, Luo20a}, which is
much lower than that of FRBs.

Additionally, although the merger models of DNS \citep{Totani13},
double black hole (BH) \citep{Abbott16} and
BH-NS \citep{Mingarelli15} are not preferably taken as the origins  of FRBs, we cannot rule out other merger and collision models,
such as the ones by the  double white dwarf (WD) \citep{Kashiyama13}, WD-NS \citep{Gu16} and BH-WD \citep{Li18}.
The reasons are that the merger rate
of WD-WD \citep{Badenes12}, WD-NS \citep{Thompson09} and BH-WD \citep{Cowperthwaite15} can be argued to satisfy with
that of FRBs, which is in several  milliseconds.%

\subsection{Supporting the magnetar origin for FRBs}
The magnetar origin models for the repeating FRBs seem to be
promising.
\Rmnum{1}. Magnetars can release soft gamma ray bursts
repeatedly, and FRB phenomena on the magnetar SGR 1935+2154
\citep{Bochenek20, CHIME20b} have been observed recently.
Meanwhile, if the non-repeaters are partly the repeaters with a
long repeating time, they may come from the highly energetic  SGRs of magnetars
with a very long reoccurrence time.
\Rmnum{2}. Some FRB repeaters are observed to have polarization characteristics
\citep{Luo20b}, which are related to the strong  magnetic
activities.

A serious problem  of the FRB-SGR association is that the birth
rate of magnetars is much higher than the event rate of FRBs.
As proposed by many researchers, the birth  rate of magnetars\footnote{\url{https://solomon.as.utexas.edu/magnetar.html}}
is about $\sim10^{-2}-10^{-3}\,yr^{-1}\,galaxy^{-1}$
\citep{Duncan92, Kouveliotou98, Gill07, Ferrario08, Gull15,
Mereghetti15, Beniamini19}, or equivalently expressed as
$\sim10^{6}-10^{7}\,Gpc^{-3}\,yr^{-1}$, which is about two orders
of magnitudes higher than the event rate of FRBs (see Table 1).
However,  if we consider a $\gamma$ factor of beaming effect and
not all hard X-ray bursts generate radio bursts \citep{Lin20},
the event rate of magnetars to release FRBs  may drop down at
least two orders of magnitudes, which can   reconcile    the
difficulty in the  event rates of FRBs and SGRs.  Here, we assume
that the beaming solid angle of the FRB is about 0.1 sr (about one
hundredth of the whole sphere).
Therefore, the rate of magnetar SGR for FRBs can be  reduced  to
$\sim10^{-5}\,yr^{-1}\,galaxy^{-1}$, which is satisfied with the
observational requirement of FRB event rate.  As remarked, in
calculating the FRB luminosity, Eq. (2) and Eq. (3) are employed to present the
luminosity in  the approximated unit solid radian. For the FRB
repeaters, if the beaming solid angle is 0.1 sr,
the FRB radio luminosity needs to be reduced by one order of magnitude.
Furthermore, there are 16
SGRs \footnote{\url{http://www.physics.mcgill.ca/~pulsar/magnetar/main.html}}
detected so far (12 confirmed, 4 candidates,
\cite{Olausen14}), and  only one case of FRB associated with
SGR has been observed \citep{Bochenek20, CHIME20b}, implying a low
rate of FRB-SGR  event, e.g.,
expressed by a  rarity factor of FRB-SGR association  of  $< 10\%$.
Thus, the reasonable event rate of FRBs can be obtained by
constraining the beaming angle and rarity factor of emitting the radio bursts. %

Similar to the FRB-SGR association in magnetar, FRBs originated from the supergiant radio pulses of the NSs could be
also possible \citep{Cordes16, Lyutikov16, Lorimer18}, although their actual occurrence rate is uncertain.
Giant pulses with the radio flux density  over ten times the normal pulses were noticed long before Crab pulsar \citep{Staelin68, Heiles70, Staelin70}.
Then, lots of giant pulses, with intensity  even  as high as thousand  times that of the average of normal pulses,  have been observed from several young pulsars  \citep{Romani01, McLaughlin03, Cordes04, Mickaliger12}.
In addition, the similarity  between the  slope of the fluence distribution for Crab giant pulses and the
repeating FRB 121102  was noticed \cite{Bera19}, which should have interesting implications on the nature of the FRB phenomena and  giant pulses.
Thus, it is meaningful to consider the association between the FRB and supergiant pulse, although the relations between magnetars
and NSs of emitting supergiant pulses are not clear yet.
Furthermore, asteroid or planetary system dominated by NS can also generate repeating FRBs in some models \citep{Dai16, Kuerban21}, which are also instructive for us to understand the physical nature of FRBs.

Finally, we  discuss the uncertainty in deriving the values
of FRB luminosity, which comes from the observational errors of
three quantities,  i.e., the flux density,   the distance inferred
by the redshift and emission solid angle of FRB.  The latter two
quantities depend on some assumptions,  which act as the dominant effects on the luminosity. For example, the uncertainty
of the emission  beaming solid angle may  cause the uncertainty
of about one order  of magnitude in determining the FRB
luminosity if we ascribe the FRB as a magnetar origin, as
discussed above. In contrast, the luminosity errors caused by the
measurement flux density will not act as a main factor. 
Therefore, our statistical analysis based on the FRB luminosity distribution implies the following idealized assumptions. 
To begin with, the one-one correlation between the redshift and DM \citep{Macquart20}.
Then, each FRB has a similar beaming solid angle of emission.
Moreover, it is assumed that all FRB measurements are accurately measured and are not affected by different telescope beams.
Apparently, more precise determinations of FRB parameters are
needed for the derivation of robust FRB luminosity properties.  

\section{Conclusions}\label{5}

To study the model constraints for FRBs, we investigate the
statistical properties of FRB luminosity  and FRB  event rates,
and find that magnetars or supergiant pulse of NSs are
favored as the origins for the repeating FRBs and GRBs should
be excluded for the non-repeaters.
In the  data selection, we only analyze the CHIME data for FRB
luminosity distribution, which avoids the uncertainty of the flux
density from different radio telescopes. 
If we assume that the current FRB distance
inferred from FRBCAT based on  the redshift and DM relation is
reliable, then the FRB luminosity distribution  should be more
consistent with the Gaussian type in logarithmic scale, which will
present a tight  constraint on the FRB origins. 
Considering  the FRB duration, radio burst luminosity and
event rate, we discuss the non-repeater and
repeater models, as listed in Table 1. For the non-repeater FRB
models,   we exclude the LGRB and SGRB origins, meanwhile
some models based on the mergers and collisions of compact objects
with the non-degenerate stars cannot be easily excluded at
present. For the repeater FRB models, we favor the magnetar or NS
supergiant pulse models. The big difference of the birth
rate of magnetar or NS from the FRB event rate can be solved by
introducing the small  beaming solid angle (0.1sr) for FRB emissions and/or
the rarity factor ($\sim$0.1) of FRB-SGR associations.

\section*{Acknowledgments}

This work is supported by the National Natural Science Foundation of China (Grant No. 11988101, No. U1938117, No. U1731238, No. 11703003 and No. 11725313),
the International Partnership Program of Chinese Academy of Sciences grant No. 114A11KYSB20160008,
the National Key R\&D Program of China No. 2016YFA0400702, and the Guizhou Provincial Science and Technology Foundation (Grant No. [2020]1Y019).
Finally, we sincerely and especially thank the anonymous referee for the meaningful comments and suggestions, which have significantly improved the quality of the paper.

\section*{Data Availability}
The data underlying this article are available in the references below:
(1) Repeating FRBs are from \citep{CHIME19a, CHIME19b};
(2) non-repeaters are from the database of FRB Catalogue (FRBCAT), available at \url{http://www.frbcat.org/};
(3) information of SGRs are from McGill Online Magnetar Catalog,
available at \url{http://www.physics.mcgill.ca/~pulsar/magnetar/main.html}.

\section*{Appendix A: different expressions for the FRB event rate}

The event rate of FRBs given by observations is often expressed in 
different units, e.g.,  $day^{-1}\,sky^{-1}$, $Gpc^{-3}\,yr^{-1}$
and $yr^{-1}\,galaxy^{-1}$ \citep{Lorimer07,Kulkarni14,Cordes19}.
To reconcile these equivalent expressions, in this appendix, we
make the  conversion  of the observation rate of
$10^{4}\,day^{-1}\,sky^{-1}$ into the units of $Gpc^{-3}\,yr^{-1}$
and $yr^{-1}galaxy^{-1}$. First, we convert it into the unit of
$Gpc^{-3}\,yr^{-1}$,
\begin{equation}
\begin{split}
10^{4}\,day^{-1}\,sky^{-1} & = \frac{10^{4}\times 365}{\frac{4}{3}\pi {\rm D_{H_0}}^3}\,Gpc^{-3}\,yr^{-1}\\
& = 1.1\times10^{4}\,Gpc^{-3}\,yr^{-1} .
\end{split}
\end{equation}
where $\rm D_{H_0}$ is Hubble distance around $4.3\,Gpc$ with Hubble constant $H_0\sim 70\,km\,s^{-1}\,Mpc^{-1}$.
Second, we convert the FRB event  rate    into  the unit of
$yr^{-1}\,galaxy^{-1}$,%
\begin{equation}
\begin{split}
10^{4}\,day^{-1}\,sky^{-1} & = \frac{10^{4}\times 365}{N_{G}}\, yr^{-1}\,galaxy^{-1}\\
&=1.8\times10^{-5}\, (\frac{{N_{G}}}{2\times 10^{11}})\, yr^{-1} \,galaxy^{-1}.
\end{split}
\end{equation}
where $N_{G}$ is the galaxy number in the observable universe,
which  ranges from $2\times 10^{11}$ \citep{Gott05} to $2\times
10^{12}$ \citep{Conselice16}, hence we take the conservative value
of $N_{G} \sim  2\times 10^{11}$ for calculation.
Meanwhile, what we need to clarify here is that the above is just a rough estimation.
This conversion is without considering the different detection limits of various radio telescopes, the curvature of the space, the luminosity scattering, etc.

\section*{Appendix B: The evaluation  of the upper limit of redshift for FRBCAT}

To evaluate the FRB distances estimated by the redshift of
the different models, we make the comparison for the upper limits
of redshift of the two models,  given by FRBCAT ($z_{CAT}$) and
those by \cite{Luo18} ($z_{Luo}$), respectively, as listed in
Table 2. The upper limit of redshift hints  the  assumptions that
the contribution of the host galaxy or surrounding material to DM
is ignored. Interestingly, we find that the values of $z_{Luo}$
are systematically  higher  than those of $z_{CAT}$ with one
extraordinary case. Especially for the well known source
FRB121102, $z_{Luo}$ (0.32) and $z_{CAT}$ (0.31) are similar, and
both are deviated from  the real value of $z_{host}$ (0.19) by
about 30 \%. Therefore, it is reasonable for us to conclude  that
the values of $z_{Luo}$ may not be necessarily better than those
of  $z_{CAT}$. So,  it is acceptable to directly quote the
inferred values of redshift and distance listed in FRBCAT.
\begin{table}
\centering \caption{Comparison of the upper limit of redshift for
FRBCAT }
\begin{tabular}{@{}lccc}
\hline
\hline
\noalign{\smallskip}
    \bf{FRB} & {$ z_{CAT}$} & {$ z_{Luo}$} & {$ (z_{Luo}-z_{CAT})/z_{Luo}$} \\
\hline
\noalign{\smallskip}
  FRB 010125 & 0.57  & 0.80  & 0.29  \\
    FRB 010621 & 0.19  & 0.48  & 0.60  \\
    FRB 010724 & 0.28  & 0.33  & 0.15  \\
    FRB 090625 & 0.72  & 0.98  & 0.27  \\
    FRB 110220 & 0.76  & 1.03  & 0.26  \\
    FRB 110523 & 0.48  & 0.67  & 0.28  \\
    FRB 110703 & 0.98  & 1.21  & 0.19  \\
    FRB 120127 & 0.43  & 0.60  & 0.28  \\
    FRB 121002 & 1.30  & 1.78  & 0.27  \\
    FRB 121102 & 0.31  & 0.32  & 0.03  \\
    FRB 130626 & 0.74  & 0.99  & 0.25  \\
    FRB 130628 & 0.35  & 0.48  & 0.27  \\
    FRB 130729 & 0.69  & 0.93  & 0.26  \\
    FRB 131104 & 0.59  & 0.63  & 0.06  \\
    FRB 140514 & 0.44  & 0.61  & 0.28  \\
    FRB 150215 & 0.57  & 0.91  & 0.37  \\
    FRB 150418 & 0.49  & 0.51  & 0.04  \\
    FRB 150610 & 1.20  & 1.66  & 0.28  \\
    FRB 150807 & 0.19  & 0.28  & 0.32  \\
    FRB 151206 & 1.50  & 2.00  & 0.25  \\
    FRB 151230 & 0.80  & 1.03  & 0.22  \\
    FRB 160102 & 2.10  & 3.10  & 0.32  \\
    FRB 160317 & 0.70  & 0.86  & 0.19  \\
    FRB 160410 & 0.18  & 0.26  & 0.31  \\
    FRB 160608 & 0.37  & 0.43  & 0.14  \\
    FRB 170107 & 0.48  & 0.66  & 0.27  \\
    FRB 170827 & 0.12  & 0.18  & 0.33  \\
    FRB 170922 & 1.20  & 1.20  & 0.00  \\
    FRB 171209 & 0.87  & 1.37  & 0.36  \\
    FRB 180309 & 0.19  & 0.27  & 0.30  \\
    FRB 180311 & 2.00  & 1.75  & -0.14  \\
\hline \hline \noalign{\smallskip}
    \end{tabular}
  \label{tab:addlabel}
\end{table}

\end{document}